\documentclass{sig-alternate} 
\usepackage{mathptmx}

\newcommand{\ignore}[1]{}
\usepackage{fancyhdr}
\usepackage[normalem]{ulem}
\usepackage[hyphens]{url}
\usepackage[keeplastbox]{flushend}
\usepackage{color}
\usepackage{subfig}
\usepackage{tikz}
\usepackage{enumitem}
\usepackage{microtype}
\usepackage[flushleft]{threeparttable}

\newcommand{\bheading}[1]{{\vspace{2pt}\noindent{\textbf{#1}}\hspace{2pt}}} 
\newcommand*\circled[1]{\tikz[baseline=(char.base)]{
            \node[shape=circle,draw,inner sep=0.7pt] (char) {#1};}}

\newlist{inparaenum}{enumerate}{2}
\setlist[inparaenum]{nosep}
\setlist[inparaenum,1]{label=\arabic*.}
\setlist[inparaenum,2]{label=(C\arabic{inparaenumi}.\arabic*),leftmargin=20pt}

\def\authnotes{1}
\newcommand{\authnote}[2]{\ifnum\authnotes=1\begin{quote}\textbf{#1 says:} #2\end{quote}\fi}
\newcommand{\fixme}[1]{\ifnum\authnotes=1\textbf{\textcolor{red}{[FIXME: #1]}}\fi}

\newenvironment{packeditemize}{
\begin{list}{$\bullet$}{
\setlength{\labelwidth}{8pt}
\setlength{\itemsep}{0pt}
\setlength{\leftmargin}{\labelwidth}
\addtolength{\leftmargin}{\labelsep}
\setlength{\parindent}{0pt}
\setlength{\listparindent}{\parindent}
\setlength{\parsep}{0pt}
\setlength{\topsep}{3pt}}}{\end{list}}

\title{Practical and Scalable Security Verification \\ of Secure Architectures} 

\numberofauthors{3}
\author{
\alignauthor
Jakub Szefer\\
       \affaddr{Yale University}\\
       \email{\normalsize{jakub.szefer@yale.edu}}
\alignauthor
Tianwei Zhang\\
       \affaddr{Princeton University}\\
       \email{\normalsize{tianweiz@alumni.princeton.edu}}
\alignauthor
Ruby B. Lee\\
       \affaddr{Princeton University}\\
       \email{\normalsize{rblee@princeton.edu}}
}

\begin{document}
\maketitle

\begin{abstract}

%In the last decade, a number of secure processor architectures have been proposed in academia, and now some are available in commercial products, such as Intel's SGX or AMD's SEV.  However, most, if not all, of the designs are not thoroughly security verified, bringing into question the security of these architectures, and systems built around them. To address this issue, 
We present a new and practical framework for security verification of secure architectures. Specifically, we break the verification task into external verification and internal verification. External verification considers the external protocols, i.e. interactions between users, compute servers, network entities, etc. 
%management servers, compute servers, trusted third parties, and the customers or users. 
Meanwhile, internal verification considers the interactions between hardware and software components within each server. This verification framework is general-purpose and can be applied to a stand-alone server, or a large-scale distributed system. 
%a distributed system or a large-scale cloud system. 
We evaluate our verification method on the CloudMonatt and HyperWall architectures as examples.
%susing both ProVerif and Murphi verification tools, as examples.

\end{abstract}

\section{Introduction}

Over the last decade, a number of secure architectures have been designed to provide security
functionalities (e.g., XOM \cite{XOM}, AEGIS \cite{suh03}, SP \cite{lee05}, Bastion \cite{champagne10}, HyperWall \cite{szefer12_asplos}, DataSafe~\cite{chen2012software}, Sanctum~\cite{costan2016sanctum}, or HDFI~\cite{song2016hdfi}).
%They all share a common feature that processor hardware is enhanced and entrusted with implementing some security functionalities.
%, such as isolation of trusted software modules from untrusted operating systems, or protecting virtual machines from untrusted hypervisors, for example.  
Ideas presented by some of these architectures have been implemented in commercial designs, such as ARM TrustZone~\cite{winter2008trusted}, Intel's SGX \cite{McAlBe:13}, AMD's SEV \cite{amd_memory_encryption}.

Once any such security architecture is designed, it is necessary to check that there are no security vulnerabilities with the design that could allow an attacker to subvert the protections. Unlike software-based solutions which may be easily patched in the field, hardware
architecture protections need to be correct from the beginning, as it is expensive and often 
not possible to update or replace them once hardware is manufactured.
%This should be done before hardware is manufactured, as it is 
%Especially, once hardware is manufactured, it is expensive and often simply not possible to update or replace it. Unlike software-based solutions which may be easily patched in the field, hardware-based protections need to be correct from the beginning since only limited hardware or firmware workarounds are possible in the field.
%the design -- the designs have to be secure 
%from the beginning, and can not be easily patched as is the case for software.
To address this issue, designers run extensive tests and simulations to make sure that the mechanisms work correctly.
Moreover, the designers perform informal security evaluation which attempts to qualitatively reason about potential 
attacks and show how the architectural mechanisms prevent them.
There is a lack, however, of a systematic methodology for verification of security architectures that can be applied to
any architecture or system in a scalable manner.

A big challenge in verifying secure architectures is that secure architectures today are usually very complex. A secure architecture is likely to consist of different types of computing servers, and the end users. All of these are connected by networks. Meanwhile each computing server consists of different layers of software and hardware components. The secure operations of the architecture include the mutual communication between servers and users across the networks, as well as interactions between hardware and software modules inside a server. So verification of complex architectures thus needs to be achieved by focusing on two key aspects: external protocols and internal interactions.

Our contribution in this paper is the definition of a general-purpose security verification framework. It has different advantages. First, it is scalable to verify complex secure architectures. The presented approach breaks down a secure architecture into smaller components for verification.
Specifically, verification of a secure architecture can be achieved effective by focusing on external verification, of the external protocols, 
and internal verification, of the internal interactions.
External protocols are used for communication between servers and users, while internal interactions are for interactions among components within each server.
We build state machines to verify the external protocols and internal interactions, thus effectively achieving verification scalability. 
%The components in these protocols and interactions are made up of various software or hardware mechanisms that themselves need to be verified. State machines and logic of these components can be verified by standard functional verification methodologies.

%(e.g. external verification treats servers as black boxes, while internal interaction takes care of checking operations within each server, see Sections \ref{sec:methodology} and \ref{sec:framework}).  

%A complex secure server today is likely to be used in a distributed scenario where there are different types of computing servers, the end users. All of these are connected by 
%is a small number of centralized management servers, a large number of computing servers, and finally the end users. All of these are connected by, possibly untrusted networks, and require mutual communication to achieve secure operation. The external protocols represent the interaction of the different parties, across untrusted networks and while being relayed by untrusted parties. 
%(e.g. untrusted cloud controller passes messages to servers which have trusted software modules running in, e.g., Intel's SGX enclaves).  
%Secure architecture verification cannot thus only focus on the secure processor itself.
% and one processor (or server), it needs to consider the whole system. Verification of complex architectures thus needs to be achieved by focusing on two additional key aspects they all share: external protocols and internal interactions. 

Second, our methodology is general-purpose and can be applied to different architectures. This method is not restricted to specific tools: designers can choose the tools they prefer to do the verification following our methodology. This achieves great practicality and granularity. We provide two case studies: verifying CloudMonatt~\cite{ZhLe:15} using a cryptographic protocol verifier ProVerif~\cite{proverif}, and verifying HyperWall \cite{szefer12_asplos, Ja:13} using a generic model checker Murphi~\cite{Murphi}. These case studies show that our methodology has been partly used to help design and enhance the secure architectures.

%Our methodology is realized in both ProVerif~\cite{proverif} and Murphi~\cite{Murphi}, showing that it is flexible and can be done in tools already familiar to designers.  ProVerif is a well-known protocol verifier, which can handle numerous interacting principals and has built-in support for cryptographic operations;  meanwhile, Murphi is a generic model checker that can model interacting principals as state machines.  They both, independently, can be used to do the verification following our methodology, giving designers flexibility to choose the tool they prefer.  This methodology has been partly used to help design and enhance the CloudMonatt~\cite{ZhLe:15,Zh17:thesis} and HyperWall \cite{szefer12_asplos, Ja:13}, further showing its practicality and granularity.  

In summary, our contributions are:
\begin{packeditemize}
\item A new, general-purpose security verification framework for secure architectures and systems.
\item A methodology to break the verification task of secure architectures and systems into external and internal
verification, which can also be done hierarchically.
\item A method to model different entities and components of such architectures as finite state machines.
\item Evaluation of the methodology on different architectures using different tools.
\end{packeditemize}

We introduce our verification methodology and framework in Section \ref{sec:methodology}. 
Using this methodology, we verify two secure architectures as case studies in Sections 
\ref{sec:verifyhyperwall} and \ref{sec:verifycloudmonatt}.
We show the verification performance in Section~\ref{sec:evalsummary}. 
%sWe discuss the evaluation results in Section \ref{sec:discussion}. 
We summarize related work in Section \ref{sec:related} and conclude in Section \ref{sec:conclusion}.

%This paper  the 
%
%, a finite-state model checker; and we show how to implement it by defining a new record data type which holds the tags associated with each value, defining functions for tag propagation, and writing our invariants to check for attacks based on the values and the associated tags.  The work presented in this chapter and the next chapter supersedes our initial exploratory work on security verification \cite{zhang12}.

\section{Verification Approach}
\label{sec:methodology}

%\subsection{Design Goals}
%\label{sec:verify_goal}

{\em The security verification of secure architectures goes beyond functional verification.} % of the mechanisms which realize the architectural features.  
%We need to worry about potential attackers who will try to intelligently manipulate inputs, outputs or any state they can access, to subvert the protections.
During design time, the threat model is specified, which lists the potential attackers and their capabilities.
The security verification methodology needs to model enough aspects of an architecture to capture all possible behaviors of these 
untrusted attackers with their capabilities, and to model their impacts on the architecture.
%their interactions with trusted components.
%A trusted component indicates that it is able to protect its integrity and confidentiality; any secrets stored by a trusted component will not leak out to attackers nor be modifiable without detection.

A secure architecture usually consists of different components (e.g., distributed nodes, software and hardware modules). 
The interactions between these components and with the external entities (e.g., remote users, networks) are very complex. 
%If attempting to perform verification of the whole system at once, we would quickly encounter the well known problem of state explosion. 
To achieve the scalability of verification, it is necessary that the verification is done on each part of the architecture, rather than on the whole architecture at once. 
Still, the verification of the sub-parts must compose into the verification of the whole architecture.

\iffalse
In summary, necessary features of the security verification methodology are the ability to:
%In summary, the high-level goals of verification are:

\begin{packeditemize}
%\item Be able to check functionality of the mechanisms which implement the different components and the protection mechanisms.
\item model both trusted and untrusted components and their interactions.
\item specify the potential attack capabilities of untrusted components, and check automatically for the possible attacks.
%\item specify security invariants for different security properties that the system should maintain.% that can be applied more broadly to different system and models.
\item verify parts of an architecture, rather than the entire system at once, to avoid the state explosion problem.
\item ensure that the verification of the parts of the system can eventually compose to the security verification of the whole secure architecture.
\end{packeditemize}

\fi

In Section~\ref{ch5:sec:methodologygoalscomponents} we propose a method of breaking the security verification of a system into smaller tasks, i.e., external verification and internal verification. In Section \ref{sec:framework} we describe the detailed steps to conduct each verification task. 
%Our approach re-uses the existing tools often used for functional verification, and applies them to the security verification task. In particular, the external protocols and internal interactions can be modelled as sets of interacting principals; and their communication can be verified using ideas from protocol verification or model checking.

%Note that the functionality of the mechanisms which implement the protections can be modelled and checked already with the existing functional verification tools. What we do propose and demonstrate is that these techniques and tools used for functional verification can be applied to the security verification task. Functional verification is thus not considered in this work.
%Given a security architecture or system, we focus only on modelling attackers and checking for 
%security vulnerabilities for external protocol or internal interaction in the secure architectures and systems. 

%%%%%%%%%%%%%%%%%%%%%%%%%%%%%%%%%%%%%%%%
\subsection{External and Internal Verification}
\label{ch5:sec:methodologygoalscomponents}

%The functional verification of the state machines ensures proper implementation of the protection mechanisms. Rather than focusing on the functional mechanisms, we look at the components that are built through these mechanisms, and their interactions. Of course, however, our work assume correct and functionally verified mechanisms. Attacks are only (potentially) present if there are untrusted components in the system. In particular, this comes into play when we consider untrusted components and their interaction with the trusted components. The untrusted components obviously have different goals (i.e. to subvert the protections) from the trusted ones (i.e. to maintain the security of the system). The security verification of the interactions ensures that the protections are upheld in spite of the potential attackers.

%\begin{figure}[t]
%  \begin{center}
%    \includegraphics[width=\linewidth]{jakub/figures/3models.png}
%    \caption[Relation between the models of the mechanisms, internal interactions and external protocols.]{\small Relation between the models of the mechanisms, internal interactions and external protocols. Each component, $C_i$, is realized through one or more mechanisms, $M_j$.}
%    \label{fig:hyperwallverif:3models}
%  \end{center}
%  \vspace{-0.5cm}
%\end{figure}

A system is composed of many components. 
%including external and internal components of a server.
Each component is realized by one or more mechanisms.
%These mechanisms can be hardware or software.  
%While internal interactions are among components only inside the local system, 
%external protocols involve components or principals outside of the local system.
%Whenever there is a change in the system, e.g. a new VM or program is started, security protections updated, attestation read, etc., there is an interaction.
We specify {\em external protocols} as the interaction of the system with distributed or
remote components, e.g. remote users, network, etc. 
There are also {\em internal interactions} which are interactions between components within a physical server or local system, e.g. 
processor, hypervisor, OS, etc.

The important aspect of the security-critical external protocols and internal interactions is that these involve untrusted principals or components, 
and hence involve potential attacks that we need to check for.
This has led us to the proposition that the components' interactions are the most important parts to verify when considering the security of the system.  
By focusing on the component interactions we have found a natural breakdown of the architecture into smaller parts.
Verifying smaller parts helps us avoid the state explosion problem.
%(see notes in Section \ref{sec:composability}).

%As shown in Figure \ref{fig:hyperwallverif:3models}, a

The security verification of the external protocols and internal interactions provides coverage of more of the system because the focus is on how 
components interface with each other, and the details of the mechanisms are abstracted away.  
A component, even a whole server,
can be treated as a blackbox during external verification -- and in turn security verified during internal verification steps.

\begin{figure}[t]
\centerline{\includegraphics[width=0.9\linewidth]{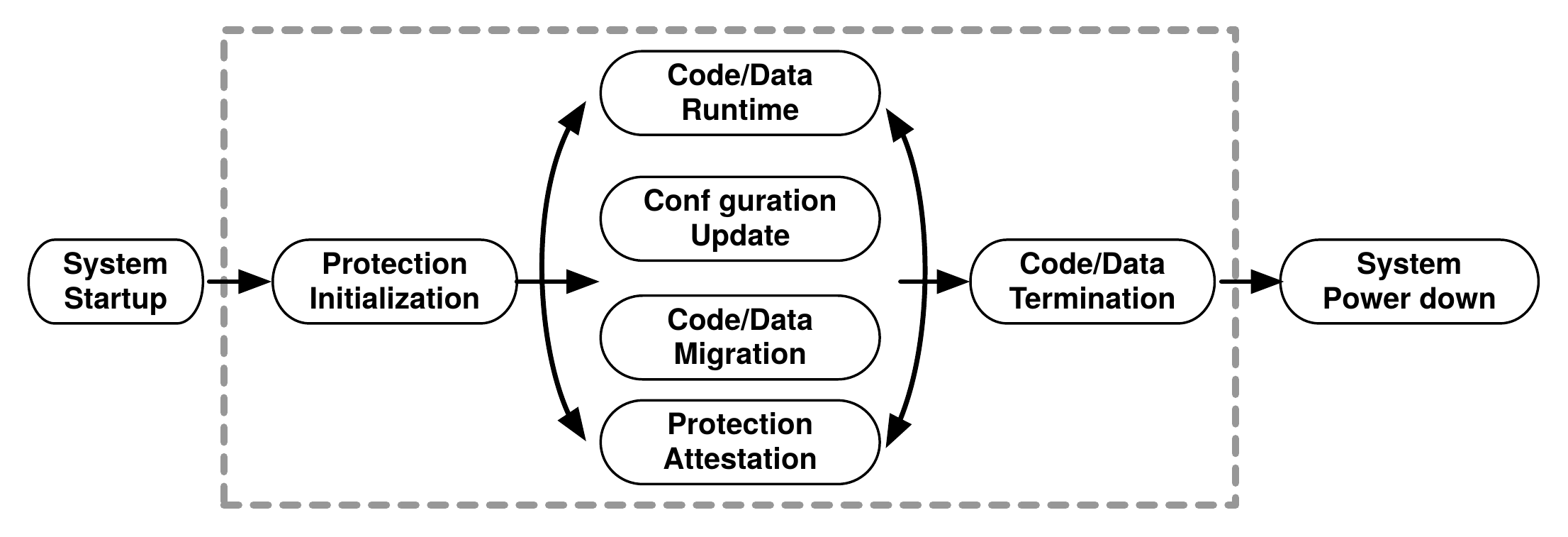} }
\vspace{-15pt}
\caption[]{\small Different execution phases of a secure architecture.} 
\label{fig:phase}
\vspace{-15pt}
\end{figure}

\bheading{Identifying protocols and interactions.}
To find the different security-sensitive interactions, we identify different execution phases of a secure architecture
or system, as shown in Figure \ref{fig:phase}.
%listed in Table \ref{table:execphases}.  
The middle six phases will be repeated many times during system runtime,
while the other two phases correspond to system startup and shutdown.
Each of the phases will have an external protocol if there is communication with the end user during that phase, 
and one or more internal interactions.  The internal interactions occur when there is an event that will cause
security-related state to be altered inside the trusted components of the architecture.
The different execution stages of a hardware secure processor architecture shown in Figure \ref{fig:phase} can be used to help identify protocols and interactions for security verification.

%\subsubsection{Secure Composition}
%\label{sec:composability}

\bheading{Secure composition.}
%Secure composition is a well known problem, which also happens in security verification.  
Given secure mechanisms or protocols, $A$ and $B$, which have been verified, it is very difficult to prove that the composition of the two is also secure.
We do not tackle the problem of formal proofs of composability in this work.  We focus on providing a sound methodology 
for the practical and scalable security verification
of individual external protocols and internal interactions.

%There has been past work on universal composability~\cite{canetti01}. More r
However, recently Protocol Composition Logic (PCL) has been 
proposed \cite{Datta07}. The composition theorems in PCL allow proofs of complex protocols to be built up from proofs of their constituent sub-protocols.
%PCL has actually been realized using Murphi, which is one of the tools that can be used to realize our methodology.  
It may be possible to build on such existing work as PCL to check the composition of the protocols and interactions which we verify.

%\subsection{Impacts of Verification on System Design}
%
%The verification can affect the system design in two ways. First, the verification process can identify the components that are necessarily trusted to guarantee the security functions and designs of the architecture. This guides the system designers to enhance the security of the architecture by applying strong protection on these trusted components. Second, it is important to note that by the time the verification process is finished, the unknown attacks will have become known as the design 
%has been iterated to fix any problems.  For example, the HyperWall architecture presented in \cite{Ja:13} includes modifications which 
%fix certain attacks that were discovered.  By examining only the final design, the verification may seem to only check for known attacks 
%(and ones which the design already protects against), but in reality the verification can find previously unknown security vulnerabilities, 
%which are then fixed in the final presented designs.

\subsection{Security Verification Framework}
\label{sec:framework}

To verify a system's protocols and operations, we first build models for the system, and identify the trusted and untrusted subjects in the system. We specify the verification goals and invariants based on the system's functionality. Then we implement the models and test through the system models. If an invariant fails in some cases, a vulnerability has been found and the design needs to be updated.

%To verify a system's external protocols and internal interactions, we first specify the verification goals and invariants
%based on the system's functionality. Invariants are the properties which should hold true for system to be secure,
%and if verification finds that an invariant is ever violated, then the is a flaw in the design that needs to be fixed.
%Then we build models for the system, and identify the trusted and untrusted subjects in the system. We can implement 
%the models and verification invariants in model checker or protocol verification tools and run the tools to test if the invariants 
%pass for every possible path through the system models from the initial state to the end state. If an invariant fails in some cases,
%it indicates a potential vulnerability (from which sample attacks can be constructed).

\subsubsection{Modeling System}
\label{ch5:essentialcomp}

\bheading{Specifying essential components.}
A designer has to enumerate the components or principals involved. For the external protocol, we treat a physical server as one component. The network component, customers, cloud provider, and (if needed) trusted third party are also explicitly included for the benefit of the external protocols which involve the remote customer connecting via a communication path to the server. For the internal interactions, we consider the hardware components (e.g., microprocessor, memory chips, co-processors) and software components (e.g., applications, hypervisor, OSes). Among the protocol or interaction participants, there are untrusted components or principals, which could be sources of potential attacks. The untrusted components or principals are the potential attackers and their capabilities need to be checked.

\bheading{Symbolic modeling.}
We adopted the symbolic modeling method \cite{Bl:12}, where the cryptographic primitives are represented by function symbols and perfect cryptography is assumed. Each component's operation can be represented as states of a state machine, and communication among the components can be represented as messages sent between the components. So we model each component as a subject. Each subject has a set of states with inputs and outputs based on the system operation. The transitions between different states are also defined by the architecture designs and protocols.

%Specifically, we first specify subjects involved in this verification procedure. A subject can be a customer or a server in the distributed system, or a hardware or software module inside a server. For the \emph{external verification} of protocols, since we treat each server as a blackbox, we model each server and the customer as a subject. For \emph{internal verification} of the interactions within a server, we need to consider the internal activities inside the server, so we model each software and hardware module involved in the system operation as a subject. Each subject has a set of states with inputs and outputs based on the system operation. The transitions between different states are also defined by the architecture designs and protocols. 

Among all the subjects, there is an {\em initiator subject} that starts the system protocol/interaction
and a {\em finisher subject} that ends the protocol/interaction; they could both be the same subject. 
This initiator subject has a {\em ``start'' state} while the finisher subject has a {\em ``commit'' state}. The verification procedure starts at the initiator's ``start'' state. At each state in each subject, it takes actions 
corresponding to the transition rules. It will exhaustively explore all possible rules and states to find all the possible paths from 
the initiator's ``start'' state to the finisher's ``commit'' state. Then we judge if the verification goals are satisfied in all of these paths. 
The system is verified to be secure if {\em there are paths from initiator's ``start'' state to finisher's ``commit'' state, and all the 
verification goals are satisfied in any of these paths.}
%, and no security invariants are violated at any time.}

%There are three main steps that need to be followed to build a representation of the architectures and implement the framework.
%
%%\begin{enumerate}[topsep=0pt,itemsep=-1ex,partopsep=1ex,parsep=1ex]
%(1) Identify external protocols and internal interactions.
%(2) Deduce essential system components.
%(3) Model the system.
%%\end{enumerate}
%%
%%\noindent 
%These tree steps are detailed in the following sub-sections.

%%%%%%%%%%%%%%%%%%%%%%%%%%%%%%%%%%%%%%%%%%%%%%%%%%%%%%%%%%%%%%%%%%%%%%%%%%%%%%%%

\subsubsection{Preconditions and Security Invariants}
The protocols and interactions are subject to constraints, the so-called preconditions.  Preconditions are closely related to the trusted computing base (TCB)
and often reflect which principals need to be in the TCB.  If a precondition is removed, the protocol or interaction may no longer be verifiable.
Ideally, during verification of a system, the minimal number of preconditions is 
determined, which can reduce the size of the Trusted Computing Base (TCB).  
One key benefit of our methodology is that it allows preconditions to be removed, (even though initially thought to be required), as verification passes with these preconditions removed.

Each protocol or interaction needs to satisfy certain security invariants -- these invariants are only verified if for all possible execution traces, the invariant is not found to be violated.
Thorough analysis of the protocols allow us to define the invariants correctly.
Often the invariant is the goal of the design so correctness is clear. 
The security invariants focus typically on confidentiality and integrity of sensitive information.
In the case of secure architectures, this sensitive information typically is: code or data executed or stored on the system, and measurements of the 
state of the system.  
%For example, the invariants are often satisfied when the architecture properly uses encryption, hashing, digital signatures and public-key 
%cryptography, with correct keys and management. 
%Our methodology could be extended to cover potential availability or authentication breaches, etc., and this can be useful future work.

\bheading{Confidentiality Validation.}
Each principal has access to various values, including ones tagged as confidential to indicate the need for confidentiality protection of that value.
%When the final state of an untrusted principal has been reached, the untrusted principal has seen all the inputs.  Now, 
The untrusted principal could try to combine all the information it has obtained in all of its states to try to break confidentiality of some of the messages (e.g. it has seen cipher text in some state, and the decryption key in another).

For each value tagged as confidential, the invariants check if any untrusted principal has access to it. If not, confidentiality of this value is maintained. Otherwise the invariants check if the value is tagged as encrypted (i.e. it has a decryption key associated with it) and the untrusted principal has access to the key. If so the untrusted principal can obtain the plaintext, thus violating confidentiality. Otherwise the confidentiality is preserved.
%that the value is also tagged as encrypted (i.e. it has a decryption key associated with it).  If a value is not encrypted, a blatant confidentiality violation exists, if this plaintext value reaches an untrusted principal. For each value tagged as confidential and encrypted, the invariants search through all the other values to see if there is a decryption key. If the untrusted principal has access to an encrypted value and the key, it can obtain the plaintext, thus violating confidentiality.
The above heuristics are consistent with our assumption of strong cryptography and that the attacker is not able to break the asymmetric or symmetric key cryptography, unless they have access to the proper key.

%These invariants are evaluated when the last state of each untrusted principal, and at the finisher's commit state, is reached. The invariants could also be evaluated earlier and the model could stop as soon as the first invariant fails.  However, this requires slightly more complicated invariants, due to extra checks for when the invariant should fire.  Given, however, that our models run quite quickly, on the order of 1 second, evaluation at the ending ``commit'' state of an untrusted principal is reasonable and easier to implement.

\bheading{Integrity Validation.}
The way we are able to check for integrity attacks is through comparing the values available to an individual trusted principal to all the values in the model.
The trusted principals have only visibility into their input values and the known-good \texttt{private} values they posses.  Meanwhile, the model has visibility into all the inputs and outputs from all the principals, and which other principals may have modified these values.  
%Thus the model has visibility into the sources of the inputs and how these sources could have modified them.  
During a run of the model, the invariants check if there is enough information in the (explicit and implicit) inputs to a trusted principal for that principal to reject any inputs that have been compromised (e.g. fabricated or replayed values). The key ideas behind the integrity checks are: (1) checking for ``known-good'' values, which can be referenced by a trusted party to validate some of the inputs, these good values need to be stored securely or come from a trusted source; (2) checking for self-consistency of values, which allows a trusted party to check the inputs and make sure they are mutually consistent.
%  \begin{itemize}
%  \item values should include a nonce for freshness, and
%  \end{itemize}
%\item 
%\end{packeditemize}

\subsubsection{Implementation and Results}

Our security verification methodology can be realized using very different tools. 
Since these are existing tools, the incremental overhead to achieve our security verification methodology is very small. Also, designers can choose the tools they are more familiar with, or that best suit their purpose. In this paper we use two verification tools ProVerif~\cite{proverif} and Murphi~\cite{Murphi} to exemplify that this is a flexible methodology. Proverif has built-in security invariant checking support which Murphi does not.  But Proverif is targeted at network protocol verification, and we have to use (repurpose) it in a clever way for checking interactions between software and hardware modules within a system. Murphi has more complete model checking facilities which enable the designer to do functional modelling and verification with the same tool as security verification. Murphi can be enhanced with security checking mechanisms as we have done, to propagate security tags for checking for integrity and confidentiality breaches.

The verification results are either 1) the protocol or interaction passes, or 2) there is some invariant that does not hold and verification fails. If verification fails, the design needs to be updated, and one has to
run the verification process again.
When verification passes, some preconditions can be removed to test if they are necessary.  Once the protocol or interaction passes with the least
number of preconditions, the verification process is completed.

In the following two sections we validate our methodology
on two types of secure architectures: a standalone server processor (HyperWall \cite{szefer12_asplos, Ja:13}). and a distributed cloud system
(CloudMonatt~\cite{ZhLe:15}).

\section{Verifying a Standalone Server}
\label{sec:verifyhyperwall}

In this section, we show how to use the above methodology to verify a secure standalone server processor. We use HyperWall \cite{szefer12_asplos, Ja:13} as an example. 

HyperWall is a secure processor architecture which aims to protect virtual machines from an untrusted hypervisor,
a predecessor to AMD's SEV extensions.  The processor hardware in HyperWall is extended with new mechanisms for managing the
memory translation and memory update so that the hypervisor is not able to compromise confidentiality and integrity of a virtual machine.
The hardware allows the hypervisor to manage the memory, but once the memory is assigned to a virtual machine, the hypervisor has no access to it.
It is scrubbed by hardware before the hypervisor can gain access again.
%Rather than use encryption, HyperWall uses isolation to ensure that memory addresses assigned to the virtual machine are off-limits to the hypervisor.
%Hypervisor can always re-claim memory, e.g., to prevent a malicious virtual machine from running forever, but the hardware scrubs 
%the memory to prevent information leakage from the virtual machine.
These protections are realized in HyperWall through extra registers and memory regions which are only accessible to the hardware, namely the TEC (Trust Evidence and Configuration) memory 
region. The TEC tables protect the memory of the guest VMs from accesses by the hypervisor 
and/or by DMA, depending on the customer's specification. Each memory region has an associated entry in the TEC tables specifying the access rights.

\begin{figure}[t]
  \begin{center}
    \includegraphics[width=0.8\linewidth]{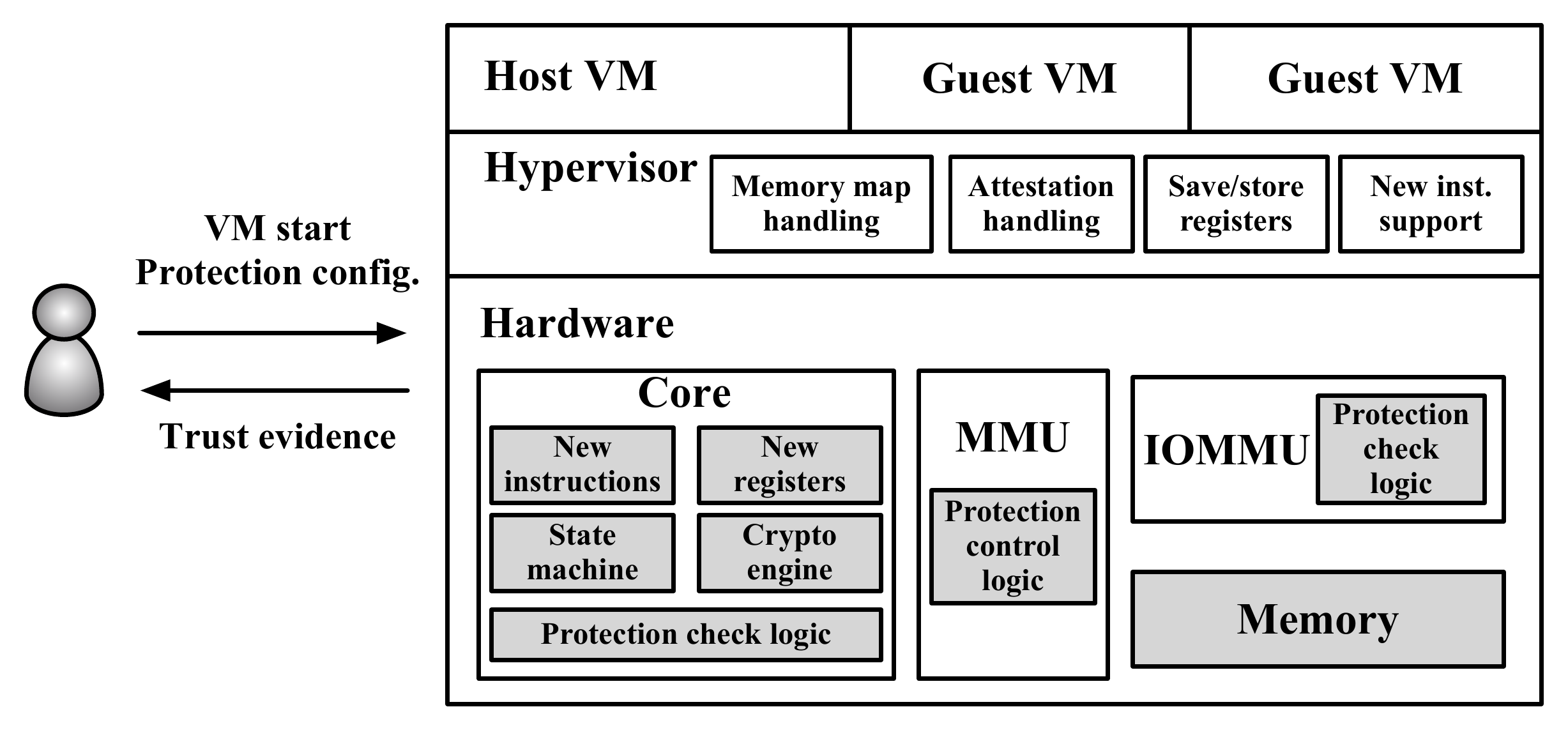}
    \vspace{-10pt}
    \caption[Hardware modification and hypervisor updates for HyperWall.]{\small
    %New hardware additions, and hypervisor updates, needed to support the HyperWall architecture.
    Architecture of HyperWall}
%HyperWall added new instructions, registers, crypto engine, core state machine, protection check logic.} 
%The MMU and I/O MMU were updated. Portion of DRAM is used to store secure data and configuration. The untrusted hypervisor and guest VMs need support for the new instructions.
    \label{fig:architectures:hardware}
  \end{center}
\vspace{-25pt}
\end{figure}

%\iffalse
%
HyperWall can be used as the cloud server in a cloud computing scenario where there is a 
remote user communicating to his or her (HyperWall) server located in the cloud possibly managed by an untrusted cloud provider.  
%In all scenarios there is also the untrusted \texttt{network} which, together with the untrusted hypervisor, is modeled as one principal in the external protocols.
HyperWall architecture is summarized in Figure \ref{fig:architectures:hardware}. Below we present verification of one external protocol and one internal interaction of HyperWall. We have further performed verification of five more HyperWall interactions, 
summarized in Section~\ref{sec:evalsummary}.

%\fi

%Hyperwall has the unusual threat model that the hypervisor is untrusted, while the virtual machine supplied by the security-conscious customer is trusted.  We illustrate our security verification methodology with  an external protocol that ensures that the Virtual Machine is launched with the VM image and protections requested by the customer on the remote server. 

%%%%%%%%%%%%%%%%%%%%%%%%%%%%%%%%%%%%%%%%%%%%%%%%%%%%%%%%%%%%%%%%

\subsection{External Protocol: VM Startup Validation}
\label{ch5:sec:verifProtVMStartup}

The security verification goal is to check if the integrity of VM image and configurations are protected during VM startup (system startup phase in Figure \ref{fig:phase}). 

%The security verification goal is to check if the untrusted network or hypervisor adversaries, between the customer and the remote server, can impact the integrity of the VM image and configurations requested (Phase 1 in Table \ref{table:execphases}).  

\begin{figure}[t]
  \includegraphics[width=0.9\linewidth]{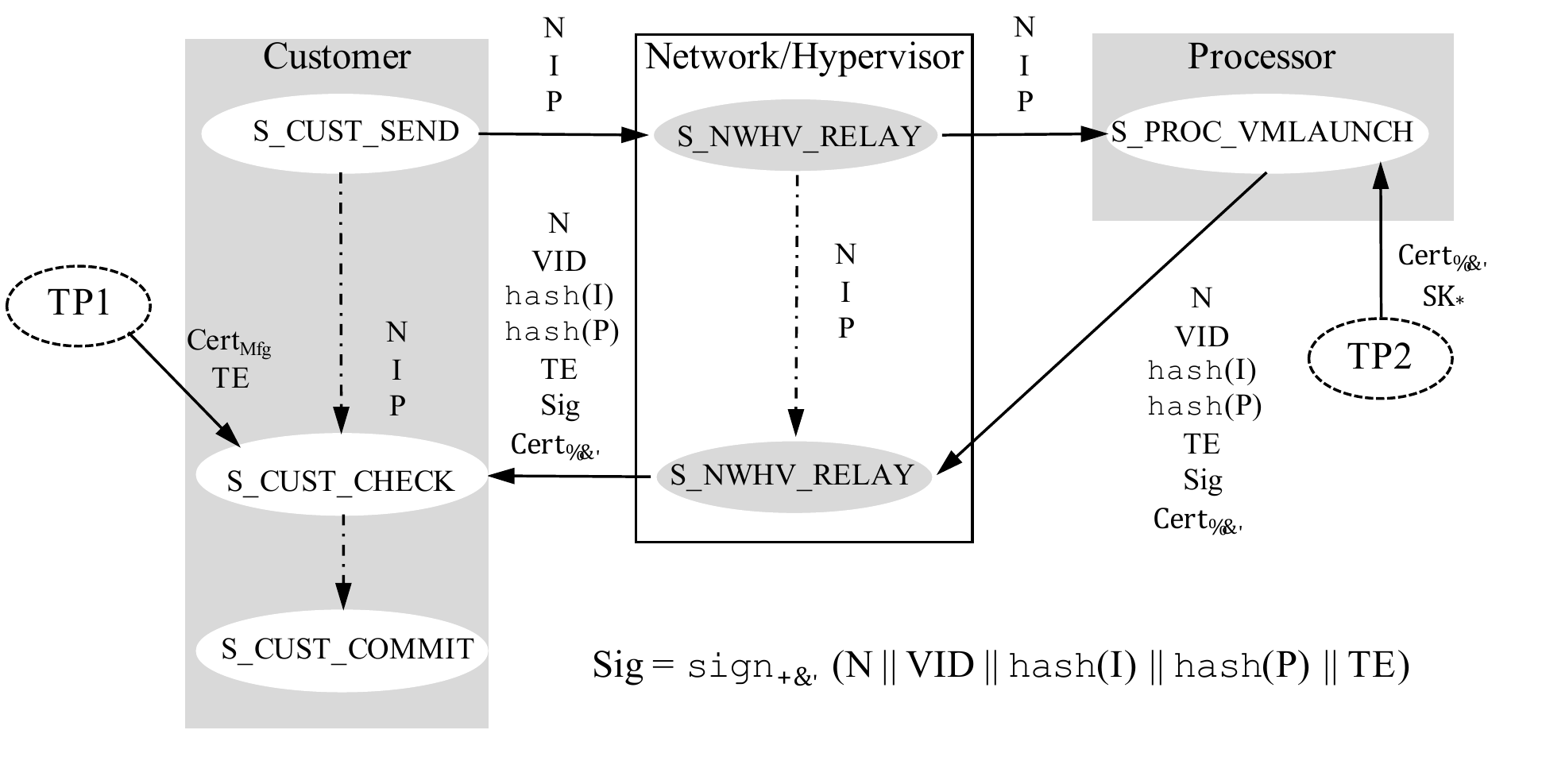}
  \vspace{-15pt}
  \caption[Model of VM Startup Protocol.]{\small Model of VM Startup Validation external protocol. \textbf{SK}$_\textbf{P}$ is the private key belonging to the
  processor, for which the customer has \textbf{Cert}$_\textbf{Mfg}$ certificate from manufacturer and the \textbf{Cert}$_{\textbf{VK}_\textbf{P}}$ certificate of the \textbf{SK}$_\textbf{P}$ sent by the \texttt{processor}, which is signed by the manufacturer.}
  \label{fig:hyperwallverif:prot_vm_startup}
\vspace{-10pt}
\end{figure}

\bheading{Modeling.}
Figure \ref{fig:hyperwallverif:prot_vm_startup} shows the external protocol with the involved components. The customer component ``starts'' VMs by specifying a nonce, \textbf{N}, the virtual machine image \textbf{I}, and the desired set of confidentiality and integrity protections for the virtual machine, \textbf{P}. This ``start VM" message is sent over the network to the hypervisor, which creates a data structure representing a VM.  
%As the network and the hypervisor are potential attackers, the values sent could be altered (fabricate different values or replay old values).  
The network and hypervisor are both untrusted and have the same attack capabilities; thus we collapse them into one component for the purpose of modeling. After the VM is prepared, the processor is invoked to start the VM, through a VM Launch instruction. The microprocessor hardware launches the VM.  It signs -- with its secret key \textbf{SK}$_{\textbf{P}}$ -- values that will define the VM: \textbf{N}, \textbf{VID} (the VM identifier assigned by the processor), \texttt{hash}(\textbf{I}), \texttt{hash}(\textbf{P}), and \textbf{TE} (the initial trust evidence where initially the number of memory access violation is zero). The five values and their signature, \textbf{Sig}, and a certificate from the hardware manufacturer with the verification key needed to check the signature, \textbf{Cert}$_{\textbf{VK}_\textbf{P}}$, are sent back to the customer. \textbf{Cert}$_{\textbf{VK}_\textbf{P}}$ is signed by the trusted vendor.

%Security verification of the protocol checks that the customer component has enough information to reject the received 
%values if any attacks on integrity have been performed.
To aid the verification, we have added two extra states to make explicit information available to the 
customer and processor.
In particular, the customer knows the certificate for the manufacturer \textbf{Cert}$_\textbf{Mfg}$ and the initial expected value of \textbf{TE}.
The processor knows the key, \textbf{SK}$_\textbf{P}$ that it uses to make the signatures.  It also has a certificate for the corresponding
public key, \textbf{VK}$_\textbf{P}$, for recipients to verify its signatures.  This information is made explicit as inputs from the two trusted 
party states, \texttt{TP1} and \texttt{TP2}.
%and it is labeled \texttt{private}.

\bheading{Security invariants.}
We identify one invariant:

\begin{enumerate}[topsep=0pt,itemsep=-1ex,partopsep=1ex,parsep=1ex,leftmargin=*,label=\protect\circled{\arabic*}]
\item The customer is able to reach the commit state with \textbf{N}, \textbf{VID}, \texttt{hash}(\textbf{I}), \texttt{hash}(\textbf{P}), \textbf{TE}, \textbf{Sig} and \textbf{Cert}$_{\textbf{VK}_\textbf{P}}$ not being compromised by the untrusted hypervisor or the untrusted network.
\end{enumerate}

%This protocol focuses on integrity of the start up values received, so the integrity invariants are used to check the protection of \textbf{NC}, \textbf{VID}, \textbf{hVM\_IMG}, \textbf{hVM\_PROT}, \textbf{TE}, \textbf{Sig} and \textbf{Cert\_VK\_hw} by having the values tagged in the model as requiring integrity protection, $INTE$.  The hypervisor is modeled as having the capability to fabricate values (triggering Prot-Inte-F-* invariants) and replay values (triggering Prot-Inte-R-* invariants).
%All eight integrity invariants are included in the verification code (Prot-Inte-F-1, Prot-Inte-F-2, Prot-Inte-F-3, Prot-Inte-F-4, Prot-Inte-R-1, Prot-Inte-R-2, Prot-Inte-R-3 and Prot-Inte-R-4 in Tables \ref{table:invariantsProtInte} and \ref{table:invariantsProtInte2}).
%The invariants will be evaluated when the customer's check state (end state) is reached to check that the customer has enough information to reject any fabricated or replayed values.

%Confidentiality invariants (Conf-1 and Conf-2) can also be included in the Murphi code.  However, since this protocol does not require confidentiality protection these invariants will not be triggered.

\bheading{Preconditions.}
We make several preconditions about the processor and cloud user and check if the above security invariants can be satisfied with these.

\begin{enumerate}[topsep=0pt,itemsep=-1ex,partopsep=1ex,parsep=1ex,leftmargin=*,label=(C\arabic*)]
  \item The processor is trusted.
  \item The processor has valid \textbf{Cert}$_{\textbf{VK}_\textbf{P}}$ and \textbf{SK}$_\textbf{P}$.
  \item The customer has valid \textbf{Cert}$_{\textbf{Mfg}}$ and \textbf{TE}.
\end{enumerate}

\bheading{Implementation.} 
We model the \texttt{Customer}, \texttt{Network}, and \texttt{Processor} in Murphi as a set of state machines.
For this protocol, we are concerned with the network or hypervisor component fabricating or replaying values as it passes them to the processor, or when it returns values 
back to the customer.  These two are collapsed into the single untrusted principal with states corresponding to two
points where this principal needs to relay the data and it could be attacked.

We extend the murphi model checker tool to propagate multiple values, for each value whose integrity must be verified: the correct value, a fabricated value and a replayed value. At the commit state, we check if the cryptography used allowed us to verify that the correct value was returned, despite transmission through the untrusted network and hypervisor.

\bheading{Results.}
The security verification passes for all possible runs and the customer can reach the commit state.
The integrity of \textbf{N}, \textbf{VID}, \texttt{hash}(\textbf{I}), \texttt{hash}(\textbf{P}), \textbf{TE}, \textbf{Sig} and \textbf{Cert}$_{\textbf{VK}_\textbf{P}}$ is protected against fabrication of values and replay of values.

Specifically, \textbf{N} and \textbf{TE} satisfy the case that there are known good values to compare against for these invariants.  For \textbf{Cert}$_{\textbf{VK}_\textbf{P}}$ there is the \textbf{Cert}$_{\textbf{Mfg}}$ that can be used to compare against it and verify it. \textbf{VID} satisfies the case that there is a signature that includes this value and a chain of certificates to verify the verification key of the signature.
\texttt{hash}(\textbf{I}) and \texttt{hash}(\textbf{P}) are hash primitives and included in the signature so they cannot be forged. The integrity of \textbf{Sig} is checked against fabrication: neither the network nor the hypervisor have access to the private signing key \textbf{SK}$_\textbf{P}$ and the customer has access to a chain of certificates that allows for him or her to verify the signature. It is also checked against replay of values: the customer can check the nonce, \textbf{N}, that he or she generated for this run of the protocol.

\subsection{Internal Interaction: VM Launch}
\label{ch5:sec:verifMechVMLaunch}

%\begin{figure}[t]
%  \begin{center}
%    \includegraphics[width=0.65\linewidth]{figures/vm_launch_addresses}
%    \caption[Review of addresses needed in looking up and setting protections.]{\small Review of addresses needed in looking up and setting protections.}
%    \label{fig:hyperwallverif:vm_launch_addresses}
%  \end{center}
%\end{figure}

We now show how to do the security verification of setting up protections for the VM's memory pages (protection initialization phase in Figure \ref{fig:phase}).

\begin{figure}[t]
  \centering
  \includegraphics[width=0.8\linewidth]{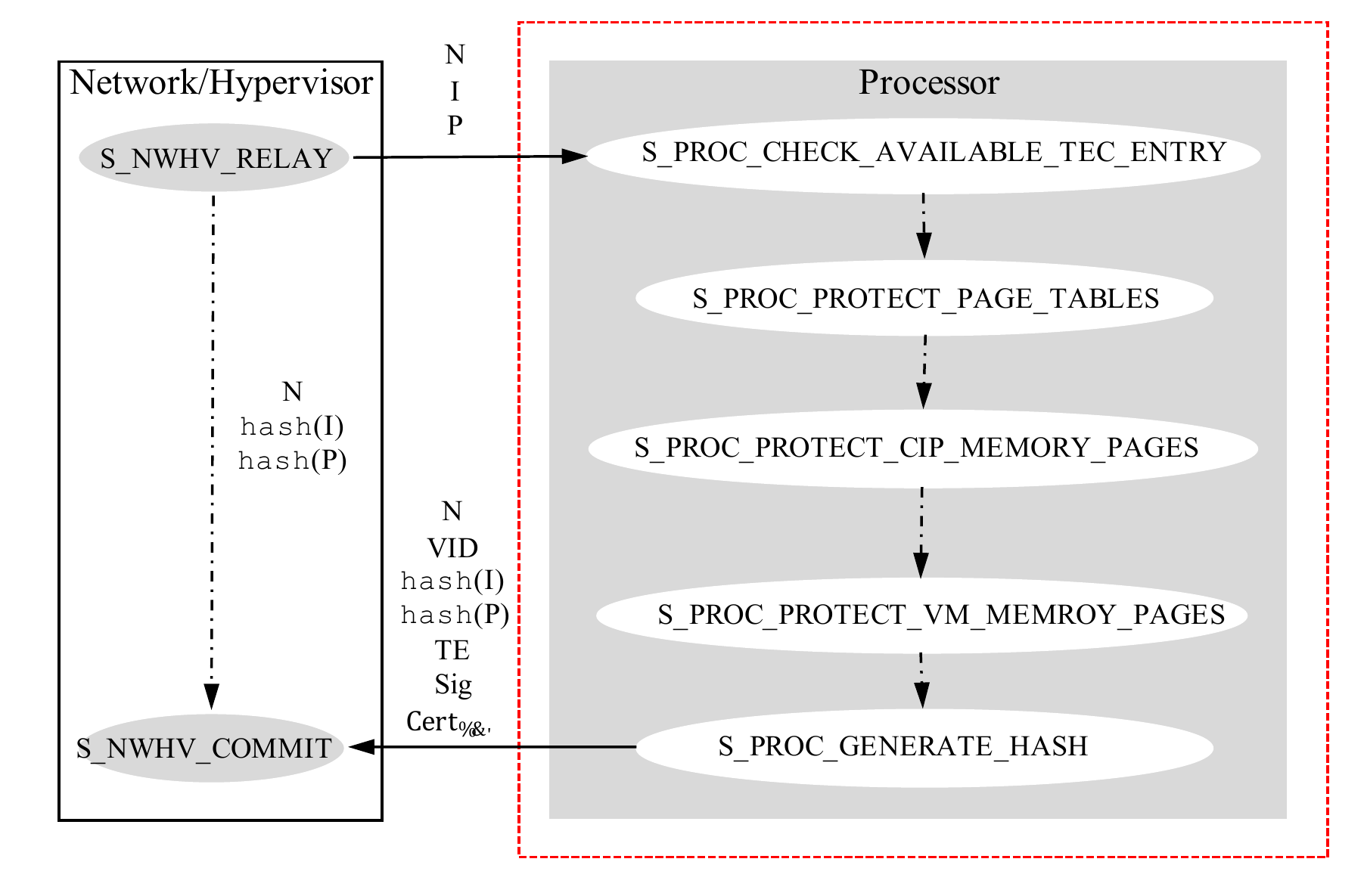}
  \vspace{-15pt}
  \caption[Model of VM Launch Mechanisms.]{\small Model of VM Launch Mechanisms. 
  \textbf{Cert}$_{\textbf{VK}_\textbf{P}}$ is the certificate of the signing key used by the \texttt{processor} in creating the signature \textbf{Sig}.
  %Note that the ``page tables'' in the figures in this chapter refer to the guest physical to machine page table mapping maintained by the hypervisor.  Also, {\em MA} refers to the machine addresses, while {\em GPA} refers to guest physical addresses.  Square boxes are states or sequences of states of the security mechanisms.
  }
    \label{fig:hyperwallverif:mech_vm_launch}
\vspace{-15pt}
\end{figure}

\bheading{Modeling.}
Figure \ref{fig:hyperwallverif:mech_vm_launch} shows the flow chart of the VM Launch mechanism. The mechanism is triggered when the hypervisor tries to start a new VM, as part of the VM startup attestation external protocol. The hypervisor sets up the VM and then executes the {\tt vm\_launch} instruction. The processor captures this instruction and atomically launches the VM with the following five operations, highlighted in Figure \ref{fig:hyperwallverif:mech_vm_launch}:

(1) The processor consults the TEC tables to find a free entry where the information about the VM will be stored. 
(2) Once a free VM entry is found, the page tables are protected. 
(3) Then the Confidentiality and Integrity Protection (CIP) tables for the VM's pages 
are protected.  
(4) The VM's pages are protected.  Each memory page is protected by denying access to the hypervisor and to DMA.
%The processor starts by reading the $VMp2m$ pointer to get the machine address of the location of the page tables.  The memory pages are then traversed so that all the memory holding the page tables can be found. Each machine page taken up by the page table data, is checked to ensure it is not in use. If there is no error, each memory page taken up by the page tables is protected by writing its corresponding entry in the CIP table (using the machine address as the index) to deny hypervisor and DMA access. 
%(3) The protection information, pre-CIP, is located, by reading the machine address from the $VMprot$ register. Each page that holds pre-CIP data is set to be protected by writing an entry in the CIP.
%(4) The memory of the VM can finally be protected.  The page table walking hardware is triggered to obtain machine address (MA) for each guest physical address (GPA) for pages that the VM is using.  For each page, the guest physical address and the assigned machine address is obtained from the page tables.  Next, the CIP tables are read (using the machine address as index) to see if the machine page is currently in use. If the page is available, the pre-CIP is read (using the guest physical address) to find the requested protections for the pages and then the CIP table is written (using the machine address again as the index) with the protections for that page.  The process repeats for all of the pages of the VM, based on the information in the page tables.
(5) Finally, the hashes of the VM image and VM protections are generated.  The page table page count is saved in the TEC table entry for the VM, and the VM is actually launched.

\bheading{Security invariants.}
We identify one invariant:

\begin{enumerate}[topsep=0pt,itemsep=-1ex,partopsep=1ex,parsep=1ex,leftmargin=*,label=\protect\circled{\arabic*}]
\item The processor needs to ensure the VM started has exactly the configuration and protection requested, and that correct hash measurements of the VM are taken.
\end{enumerate}

\bheading{Preconditions.}
We require several preconditions about the processor, these are a subset of the preconditions needed by the prior external protocol.

\begin{enumerate}[topsep=0pt,itemsep=-1ex,partopsep=1ex,parsep=1ex,leftmargin=*,label=(C\arabic*)]
  \item The processor is trusted.
  \item The processor has valid \textbf{Cert}$_{\textbf{VK}_\textbf{P}}$ and \textbf{SK}$_\textbf{P}$.
\end{enumerate}

\bheading{Implementation.}
As above, we model the untrusted network and untrusted hypervisor as a single entity, with the capability to fabricate values and replay values. The processor is trusted based on our preconditions. We use Murphi to model the processor as a state machine. %following the launch mechanism as described above. 
The \texttt{Processor} needs to ensure the integrity of the start up values received when a request to launch a VM is received: \textbf{N}, \textbf{VID}, \texttt{hash}(\textbf{I}), \texttt{hash}(\textbf{P}), \textbf{TE}, \textbf{Sig} and \textbf{Cert}$_{\textbf{VK}_\textbf{P}}$. We tag these values as requiring integrity protection and check if these values are fabricated or replayed when the protocol reaches the commit state.

\bheading{Results.}
This protocol focuses on integrity of the start up values received:
\textbf{N}, \textbf{VID}, \texttt{hash}(\textbf{I}), \texttt{hash}(\textbf{P}), \textbf{TE}, \textbf{Sig} and \textbf{Cert}$_{\textbf{VK}_\textbf{P}}$. The model keeps track of whether the reads or writes to protection tables
were accessed only by the trusted hardware. Our verification results indicate that the processor will correctly conduct the above five steps, and generate the correct hash measurements at the commit state. 

\subsection{Security Discussion}

\bheading{Coverage.}
%The above security verification shows that the main HyperWall (after the small corrections uncovered during the verification) is secure.
In addition to the two protocols shown above, five other protocols or interactions were verified, as listed in Table~\ref{table:verifProtSummary}. The protocols and interactions verified cover the execution phases from Figure \ref{fig:phase}, except for VM migration. 
The methodology facilitates a ``design for security'' approach where architects can validate
individual protocols and interactions at the design phase.  

\bheading{Impact.}
The verification effort uncovered two flaws in the original design~\cite{szefer12_asplos}, and later fixed in \cite{Ja:13}.  The first was a replay attack in the VM Suspend and Resume protocol.
The original design \cite{szefer12_asplos} included a nonce to prevent replay attacks. However, when modeling the internal interaction due to VM Suspend \& Resume, 
the verification of the model failed, pointing out that the ``nonce'' value was not updated during the suspend and resume operation as originally assumed, 
thus not providing replay protection. 
%The problem was fixed by adding
%Having discovered the problem, it was suggested that it can be solved as follows: instead of using registers to store the nonce, a new field in the TEC (Trust Evidence and Configuration) tables was defined and there 
%a dedicated nonce for VM Suspend \& Resume was stored.  
%
A related problem was discovered about the trust evidence data, previously also only stored in registers.  
Stale trust evidence data could have 
been sent back to the customer, by a compromised hypervisor. 

\section{Verifying A Distributed System}
\label{sec:verifycloudmonatt}

\begin{figure}[t]
\centerline{ \mbox{\includegraphics[width=0.9\linewidth]{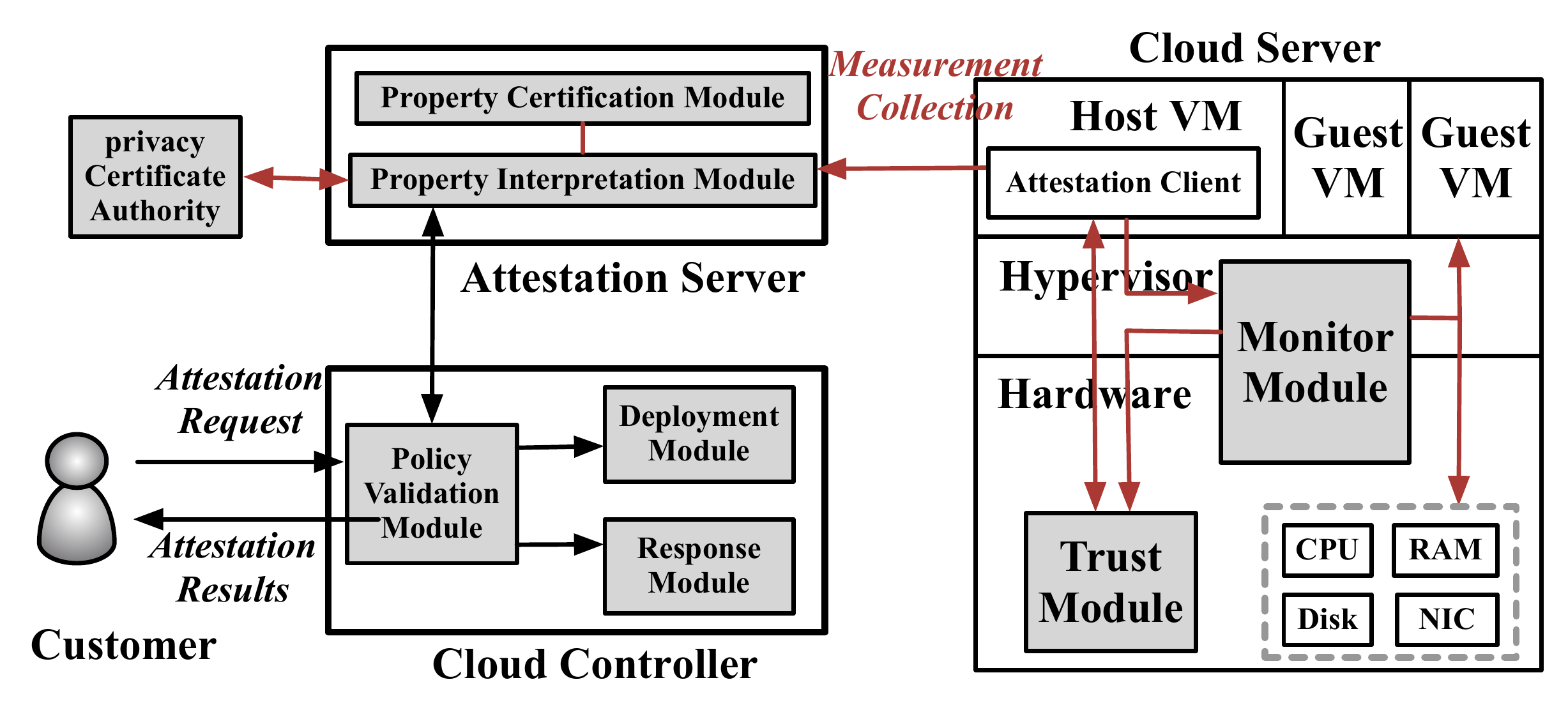}} }
\vspace{-10pt}
\caption{\small Architecture of CloudMonatt.}
\label{fig:overview}
\vspace{-15pt}
\end{figure}

CloudMonatt \cite{ZhLe:15}
is a flexible distributed cloud architecture to monitor and attest the security 
health of customers' VMs in the cloud. Figure \ref{fig:overview} shows the architecture overview of CloudMonatt. It involves four entities: the customer, the Cloud Controller, the Attestation Server and the cloud server. 
The Cloud Controller acts as the cloud manager, responsible for taking VM requests and servicing them for each customer. The Attestation Server acts as the attestation requester and appraiser, to collect the security measurements from the VM, interpret the measurements and make attestation decisions. 
The Cloud Server 
%includes an \texttt{Attestation Client} for taking requests from the Attestation Server. It
has a \texttt{Monitor Module} which contains different types of monitors to provide comprehensive and rich security measurements. It has a \texttt{Trust Module} responsible for server authentication, secure measurement storage and crypto operations.
%(e.g., encryption, signatures generation).  
%These modules use one of the existing secure processor architectures, such as Bastion or Intel SGX.

We now show how our security verification methodology can be used to verify the main attestation protocol of CloudMonatt.  
We also show how this methodology can help to narrow down the number of trusted components 
needed in the trusted computing base for this distributed system. 

\begin{figure*}[t]
\centerline{ \includegraphics[width=0.9\linewidth]{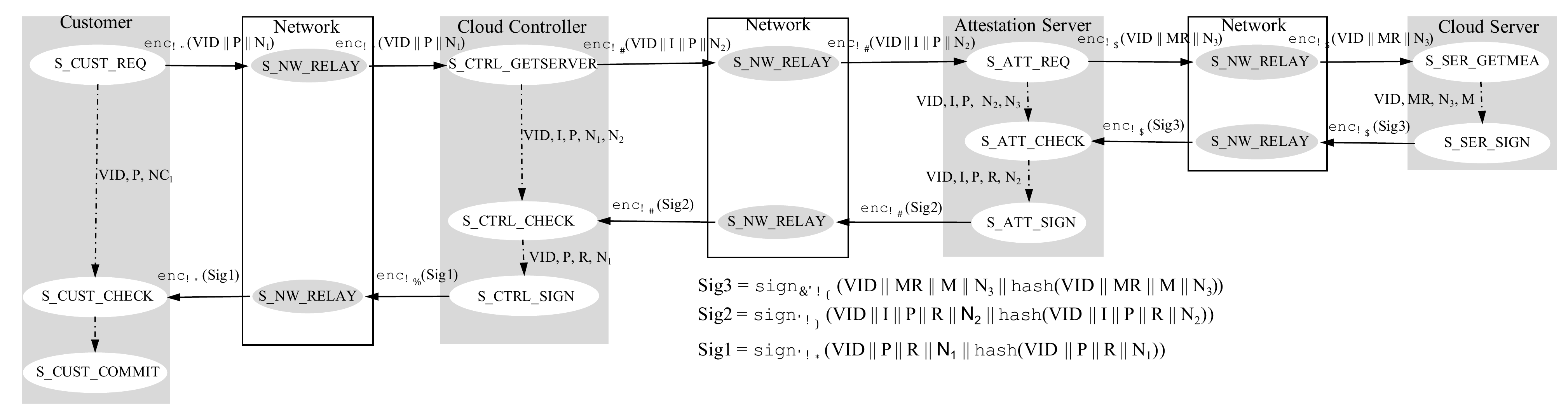} }
\vspace{-15pt}
\caption[The external protocol for attestation in \emph{CloudMonatt}.]{\small The external protocol in \emph{CloudMonatt}. \textbf{SK}$_\textbf{C}$, \textbf{SK}$_\textbf{A}$ and \textbf{ASK}$_\textbf{S}$ are the private signing keys of the Cloud Controller, the Attestation Server and the cloud server, respectively. \textbf{K}$_\textbf{X}$, \textbf{K}$_\textbf{Y}$ and \textbf{K}$_\textbf{Z}$ are symmetric keys between the customer and the Cloud Controller, between the Cloud Controller and the Attestation Server, and between the Attestation Server and the cloud server, respectively.} 
\label{fig:attest:external_protocol}
\vspace{-10pt}
\end{figure*}

\subsection{External Protocol: Cloud Attestation}
\label{sec:attest:external_verif}

Cloud attestation is the {\em procedure of making unforgeable claims about the security conditions of customers' VMs based on the evidence supplied by the host server}. %We are only interested in 
We verify that the requested report is not tampered with in CloudMonatt architecture, and hence the integrity of the end-to-end attestation is achieved (protection attestation phase in Figure \ref{fig:phase}).
%Our verification focuses on the unforgeability of the main attestation protocol of the CloudMonatt architecture. %We do not consider other activities within the cloud system.
%This requires showing that the requested attestation report has not been tampered with, and hence the integrity of the end-to-end attestation is achieved.

\bheading{Modeling.}
We model each entity involved in this distributed system as an interacting state machine, as shown in Figure \ref{fig:attest:external_protocol}. 
%Each subject is made up of some states. The customer is the initiator as well as the finisher subject. 
The whole process starts from the customer, who sends to the Cloud Controller the attestation request including the VM identifier, \textbf{VID}, and the security properties, \textbf{P}. Then the Cloud Controller 
%discovers the host cloud server, and 
forwards the request to the Attestation Server, with the host servers identifier, \textbf{I}. The Attestation 
Server sends \textbf{MR}, the request of necessary measurement, to the host server.
%identifies the necessary measurements and sends the measurement request \textbf{MR} to the host cloud server. 
The cloud server collects the required measurements \textbf{M}, hashes and signs the measurements, and sends them back to the Attestation Server.
The Attestation Server checks the received message and, if correct, generates the attestation report \textbf{R} based on \textbf{M} and \textbf{P}. 
%signature, the hash value and the nonce: if this check is correct, it interprets the measurements and the property, and generates the attestation report, \textbf{R}. 
Then the Attestation Server signs the report and transmits it to the Cloud 
Controller. The Cloud Controller checks the message and, if correct, hashes and signs the report, and sends it to the customer. The customer ends the attestation session if the he finds the report is correct.
%After receiving the report, the Cloud Controller checks the signature, the hash value and the nonce. If the check is correct, the Cloud Controller hashes and signs the report, and sends the report to the customer. If the customer finds the encrypted signature of the report is correct, it ends the attestation request session.

\bheading{Security invariants.}
We identify one invariant:

\begin{enumerate}[topsep=0pt,itemsep=-1ex,partopsep=1ex,parsep=1ex,leftmargin=*,label=\protect\circled{\arabic*}]
%\item The attestation report \textbf{R} the Cloud Controller receives is indeed the one for VM \textbf{VID} with property \textbf{P}, specified by the customer.
\item The attestation report \textbf{R} the customer receives is indeed the one for \textbf{VID} with \textbf{P}, specified by the customer.
\end{enumerate}

%\noindent Invariant \circled{1} is to ensure the Cloud Controller gets the correct attestation reports. 

\bheading{Preconditions.}
%We require 
Initially, we specify several preconditions and check if the above invariant can be satisfied under these preconditions.
Later, we verify each of these preconditions.

\begin{enumerate}[topsep=0pt,itemsep=-1ex,partopsep=1ex,parsep=1ex,leftmargin=*,label=(C\arabic*)]
  \item The cloud server is trusted. %It will collect and sign correct measurements;
  \item The Attestation Server is trusted. %It will process the received measurements and generate the reports correctly;
  \item The Cloud Controller is trusted. %It will process the VM health reports correctly.
\end{enumerate}

%Here a ``trusted'' server means it will strictly follow the operations indicated in our protocol. For instance, a trusted cloud server will collect and sign correct measurements; a trusted Attestation Server will process the received measurements and generate the reports correctly; a trusted Cloud Controller will process the VM health reports correctly. In addition, a trusted server will keep its secrets from attackers. 
%However, the network in our threat model is not trusted. The adversary can eavesdrop as well as falsify any of the messages transmitted between different servers. 

\bheading{Implementation.} 
We model the external protocol in ProVerif. Specifically, we declare each subject as a process. 
%Inside the process we model the operations of state machines shown in Figure \ref{fig:attest:external_protocol}. 
Each process keeps some variables. If the subject is trusted, we denote these variables as \texttt{private}, not accessible by the attacker. Otherwise the variables are assumed public.
%then the attacker cannot get these secrets, and we use the keyword \texttt{private} to denote these variables. Otherwise the variables are assumed public.
%To model the network activities in this distributed system, 
We declare a \texttt{network} connected between each pair of subjects, to represent the untrusted communication channels. These channels are under full control of the network-level adversaries, who can eavesdrop or modify any messages. We use the cryptographic primitives from ProVerif to model the public key infrastructure for digital certificate, authentication and key exchange. 
Then we model the attestation process for an unbounded number of sessions, and check if the adversary can compromise the integrity of the report in any session.
%i.e., the customer keeps sending attestation requests to the cloud system and receiving the reports. ProVerif can check if the adversary can compromise the integrity of the report in any attestation session, and display the attack execution trace if a vulnerability is found. 

\begin{figure}[t]
\centerline{ \includegraphics[width=0.9\linewidth]{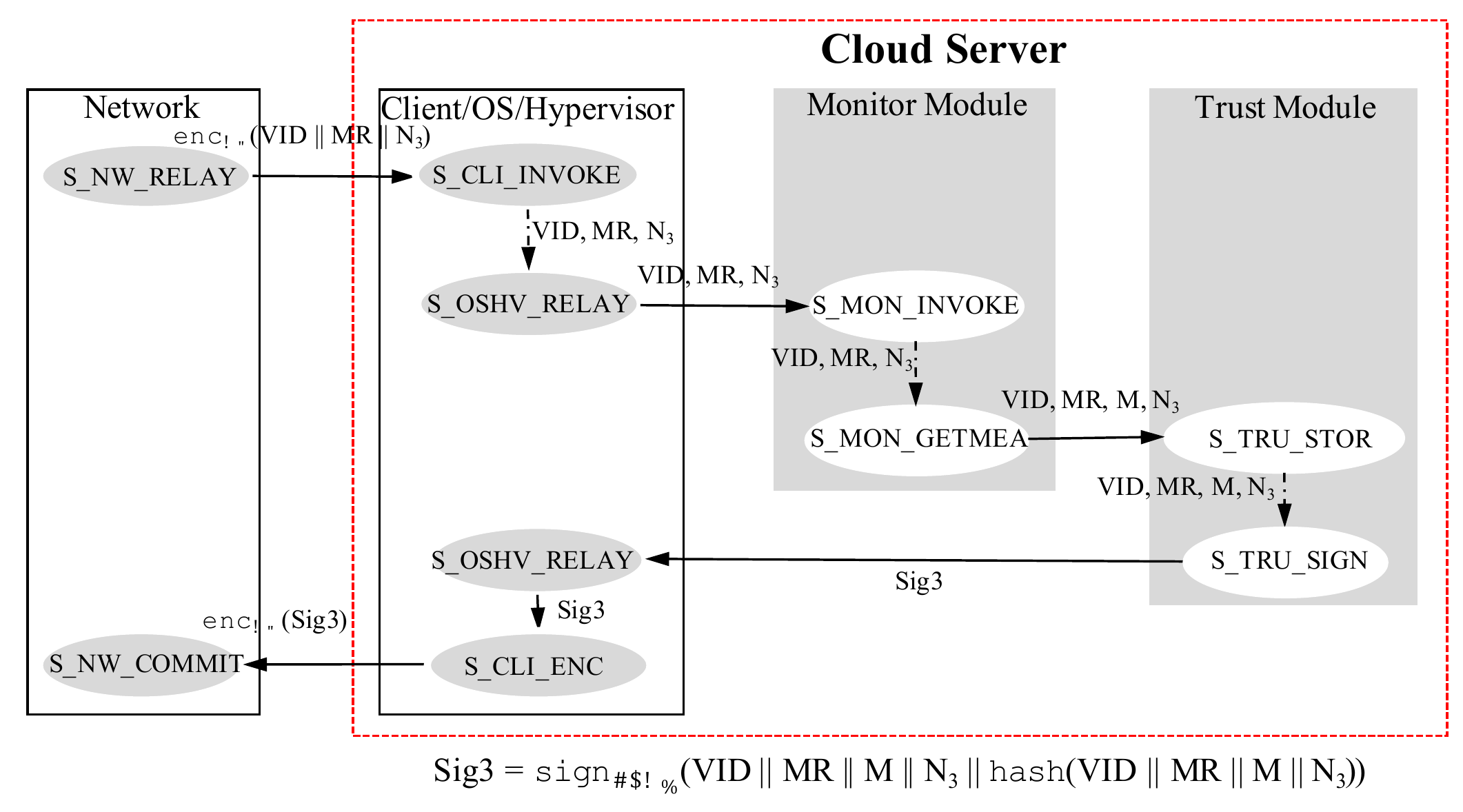} }
\vspace{-15pt}
\caption{\small Internal interactions in the cloud server. \textbf{K}$_\textbf{Z}$ is the symmetric key known to the Attestation Server and the cloud server. \textbf{ASK}$_\textbf{S}$ is the private signing key of the cloud server.} 
\label{fig:attest:internal_protocol_server}
\vspace{-10pt}
\end{figure}

We use ProVerif's \emph{reachability proof} functionality to verify the integrity of a message. 
%ProVerif allows us to define an event \textbf{E} inside a process at one state, which specifies some conditions. Then we can check if this event will happen when the protocol proceeds using the query statement: ``\texttt{query} \texttt{event}(\textbf{E})''. ProVerif can enumerate all the possible traces and check if this event is reachable in some cases. 
%If so, this query statement returns true as well as the trace that reaches the event. 
%Otherwise the statement returns false. 
Specifically, we define a function $\mathbb{R}$(\textbf{VID}, \textbf{P}) to denote the correct report of VM \textbf{VID} for property \textbf{P}. At the customer's state ``S$\_$CUST$\_$COMMIT'', the customer receives the report \textbf{R}, and we check if the statement \textbf{R} = $\mathbb{R}$(\textbf{VID}, \textbf{P}) is always true. We use the statement ``\texttt{query} \texttt{event}(\textbf{R}$\neq$$\mathbb{R}$(\textbf{VID}, \textbf{P}))" to check the negative scenario: an integrity breach has occurred. If this query statement is false, the attacker has no means to change the message \textbf{R} without being observed by the customer and the integrity of \textbf{R} holds. 

\bheading{Results.} 
First, ProVerif shows the security invariant \circled{1} is satisfied under the preconditions (C1) -- (C3). The network-level adversaries cannot compromise the integrity of the messages without being observed, as all the messages are cryptographically protected. 
%Even though the network-level adversaries can take control of all the network channels between each server, they cannot compromise the integrity of the messages without being observed, since all the messages are hashed, signed and encrypted before being sent to the network. 
Second, ProVerif shows that preconditions (C1) -- (C3) are necessary to keep the invariants correct, and missing any precondition can lead to violations of the invariant. An untrusted cloud server can counterfeit wrong measurements, making the customer receives wrong attestation report generated from the measurements. An untrusted Attestation Server can generate wrong attestation report for the customer. An untrusted Cloud Controller can modify the attestation reports before sending to the customer. 

%if the cloud server is not trusted, then the server-level adversary can counterfeit wrong measurements, causing the Attestation Server to make wrong attestation decisions, and pass them to the Cloud Controller and the customer. If the Attestation Server is not trusted, then it can generate wrong attestation reports for the customer and the Cloud Controller. If the Cloud Controller is untrusted, it can modify the attestation reports before sending to the customer. In all three cases, invariant \circled{1} is not satisfied.

%We now know that to satisfy the external verification goals, we need to assume the correctness of the preconditions, i.e., trusting the data processing in the Cloud Controller, the Attestation Server and the cloud server. 

\subsection{Internal Interaction: Evidence Collection}
\label{sec:attest:internal_verif}

Placing the entire server into the TCB would require stronger security protection, which is expensive and difficult to achieve. So, we conduct \emph{internal verification} to identify the necessary components inside the servers that need to be trusted. We verify the evidence collection process in the cloud server (protection attestation phase in Figure \ref{fig:phase}).

%We show the most complicated case: internal verification of the cloud server. The verification of the Cloud Controller and Attestation Server can be done in a similar way (results in Section \ref{sec:evalsummary}).

\bheading{Modeling.}
We model the key components inside a cloud server as state machines (Figure \ref{fig:attest:internal_protocol_server}). We also include the untrusted network as the initiator and finisher subject in the internal protocol to interact with the server. The whole process starts when the network passed the encrypted measurement request to the server. The \texttt{Attestation Client} processes the request and passes it to the \texttt{Monitor Module}. The \texttt{Monitor Module} collects the correct measurements, and then stores the measurements together with other related information in the \texttt{Trust Module}. The \texttt{Trust Module} calculates the hash and signature using its private attestation key. Then the signature is encrypted by the \texttt{Attestation Client} and sent out. The network goes to the commit state when it receives the encrypted measurement.

%The \texttt{Monitor Kernel} inside the \texttt{Monitor Module} figures out the corresponding monitor tool and invokes it to collect the correct measurements. Then it stores the measurements together with other related information in the \texttt{Trust Evidence Registers}. Then the \texttt{Crypto Engine} in the \texttt{Trust Module} retrieves the measurements, calculates the quote and signs it using the \texttt{Attestation Key}. Then the signature is encrypted by the \texttt{Attestation Client} and sent to the Attestation Server. The Attestation Server will check the hash and signature. It goes to the commit state if the check succeeds. 

\bheading{Security invariants.} We identify one invariant:

\begin{enumerate}[topsep=0pt,itemsep=-1ex,partopsep=1ex,parsep=1ex,leftmargin=*,label=\protect\circled{\arabic*}]
\item The cloud server needs to ensure that the correct measurement \textbf{M} are taken for VM \textbf{VID} with request \textbf{MR}.
\end{enumerate}

\bheading{Preconditions.}
We identify a set of possible preconditions.
% to satisfy the above invariant. 
%We classify these preconditions as different modules and inter-module communications. 
%We check the necessity and sufficiency of these preconditions for guaranteeing the integrity of measurements taken from the cloud server.  
%Later, we will try to do the security verification while eliminating each of the preconditions, to see which ones are really necessary for the TCB.

\begin{enumerate}[topsep=0pt,itemsep=-1ex,partopsep=1ex,parsep=1ex,leftmargin=*,label=(C\arabic*)]
%  \item The \texttt{Attestation Client} is trusted.%, i.e., it can maintain the integrity of the attestation requests and measurements.
  \item The \texttt{Monitor Module} is trusted.%, i.e., it can invoke the correct measurement and maintain the integrity of the measurements.
  \item The \texttt{Trust Module} is trusted.%, i.e., it can correctly calculate the hash quote and sign the measurements, maintain the integrity of measurements, confidentiality and integrity of the attestation keys. 
%  \item The channel between the \texttt{Attestation Client} and the \texttt{Monitor Module} is trusted.
%  \item The channel between the \texttt{Attestation Client} and the \texttt{Trust Module} is trusted.
  \item The channel between the \texttt{Monitor Module} and the \texttt{Trust Module} is trusted.
\end{enumerate}

\iffalse

\begin{inparaenum}
  \item \texttt{Attestation Client}:
  \begin{inparaenum}
    \item this module is trusted.
  \end{inparaenum}
  \item \texttt{Monitor Module}:
  \begin{inparaenum}
    \item the \texttt{Monitor Kernel} is trusted.
    \item the \texttt{Monitor Tools} are trusted.
    \item the channel between the \texttt{Monitor Kernel} and the \texttt{Monitor Tools} is trusted.
  \end{inparaenum}
  \item \texttt{Trust Module}:
  \begin{inparaenum}
    \item the \texttt{Crypto Engine} is trusted.
    \item the \texttt{Trust Evidence Registers} are trusted.
    \item the \texttt{Attestation Key} is securely stored.
    \item the channel between the \texttt{Attestation Key} and the \texttt{Crypto Engine} is trusted.
    \item The channel between the \texttt{Crypto Engine} and the \texttt{Trust Evidence Registers} is trusted.
  \end{inparaenum}
  \item Inter-module communication:
  \begin{inparaenum}
    \item the channel between the \texttt{Attestation Client} and the \texttt{Monitor Kernel} is trusted.
    \item The channel between the \texttt{Attestation Client} and the \texttt{Crypto Engine} is trusted.
    \item The channel between the \texttt{Monitor Kernel} and the \texttt{Trust Evidence Registers} is trusted.
  \end{inparaenum}
\end{inparaenum}

\fi

\bheading{Implementation.} 
%We can model the server system as a network system, and verify the server in a similar way as the network protocol verification. Specifically, 
We model a software or hardware component as a process. Each component keeps some variables and operates as a state machine. If 
one component is in the TCB, then its variables will be declared as \texttt{private}. Otherwise its variables are public to attackers. If two modules are linked by an untrusted channel, then we declare a public \texttt{network} between these two components. Otherwise we combine the two component into one process so that they can exchange messages securely.

%If the attacker has the privilege to control the communication between two components, then we declare a public \texttt{network} for these two components. Otherwise if two modules are linked by one channel that is trusted, then we combine the two processes into one process so that the two modules can exchange messages securely. % without being compromised by third party attackers. 

We also use ProVerif's reachability proof functionality to verify the integrity of measurement \textbf{M}. When the network reaches state ``S$\_$NW$\_$COMMIT'', we denote the measurement inside the encrypted message as \textbf{M}. We also defines a function $\mathbb{M}$(\textbf{VID}, \textbf{MR}), which gives the correct measurement of VM \textbf{VID} for the measurement request \textbf{MR}. Then we check if the statement \textbf{M} = $\mathbb{M}$(\textbf{VID}, \textbf{MR}) is always true at the commit state. We use the statement ``\texttt{query} \texttt{event}(\textbf{M}$\neq$$\mathbb{M}$(\textbf{VID}, \textbf{MR}))" to discover potential integrity breach. If this statement is false, the attacker has no means to change \textbf{M} without being observed by the customer and the integrity of \textbf{M} holds.

%We model all the steps in Figure \ref{fig:attest:internal_protocol_server} for an unbounded number of sessions, i.e., the Attestation Server keeps sending measurement requests to the cloud server and receiving the results. ProVerif enumerates all the possible states during the infinite sessions and checks if the property is maintained. 

\bheading{Results.} 
We verify that it is sufficient and necessary to keep the security invariant with these preconditions, when the network, OS and hypervisor is untrusted. Missing any prediction can lead to invariant violation: an untrusted \texttt{Monitor Module} can collect wrong measurements \textbf{M} and store them into the \texttt{Trust Module}; an untrusted \texttt{Trust Module} can generate a fake signature over any measurements using the signing key \textbf{ASK}$_\textbf{S}$; an untrusted channel between the \texttt{Monitor Module} and \texttt{Trust Module} gives the adversary a chance to modify the measurements without being detected.

\subsection{Security Discussion}

\bheading{Coverage.}
We show the main CloudMonatt attestation protocol is secure, i.e., correct and unforgeable. We show the evidence collection process in the cloud server is secure. We also verified the property interpretation process in the Attestation Server and the health checking process in the Cloud Controller in the same way as we showed for the Cloud Server. This completes the end-to-end security verification of the protection attestation phase in CloudMonatt.
%CloudMonatt attestation protocol and mechanisms.

%The security verification in Sections \ref{sec:attest:external_verif} and \ref{sec:attest:internal_verif} shows that the main CloudMonatt attestation protocol is secure, i.e., correct and unforgeable.  The attestation report requested cannot be tampered with, undetected. The cryptography, nonces and signing keys used are sufficient to detect any integrity breaches.  The security invariants did not require confidentiality, so the encryption of the communications between the servers with the secret keys is additional security for confidentiality that is optional. Confidentiality is not a fundamental requirement in this scenario, but has also been provided.  To complete the security verification coverage, we need to do the security verification of the preconditions C2 (\texttt{Monitor Module}), C3  (\texttt{Trust Module}) and C6 (the link between them).  We described how this can be done in Section \ref{sec:attest:internal_verif}, but do not detail it in this paper. Also, the security verification of the Attestation Server and the Cloud Controller must be done, in the same way as we showed for the Cloud Server. This would complete the end-to-end security verification of the CloudMonatt attestation protocol and mechanisms.

\bheading{Impact.}
One of the most interesting results of security verification is to show how we can enhance the security of the architecture during design. In CloudMonatt, it showed that only the Monitor Module and Trust Module of a cloud server should be included in the TCB. Normally, third party customers (at guest VM privilege) has no capability to subvert the security functions provided by these two modules (at the hypervisor privilege). To defeat attacks (e.g., privilege escalation) caused by the vulnerabilities of the original system, secure enclaves can be used to protect the execution environment of the Monitor Module and Trust Module, leveraging mechanisms provided by Bastion \cite{ChLe:10}.

\section{Verification Evaluation}
\label{sec:evalsummary}

In addition to the protocols presented in this paper, we have also verify five more for HyperWall and two more for CloudMonatt. For HyperWall, we use CMurphi 5.4.4 and the models were run with options {\tt -tv -ndl -m1000}. The {\tt -tv} writes a violating trace (if an invariant fails), and the {\tt -ndl} disables the checking for deadlock states. For CloudMonatt we use ProVerif 1.88 with default options.
% (if a model seems to run for an extended period of time, deadlock checking can be re-enabled to ensure that the model is not stuck). 

The collected results for HyperWall in Table \ref{table:verifProtSummary} and for CloudMonatt in Table \ref{table:verifProtSummary1}.
%In the tables we list what is being modeled, followed by the lines of code and the run time.
%The lines of code does include some comments which are very helpful for understanding the 
%verification. 
The verification process is iterative, where the ProVerif or Murphi files may be 
updated many times, thus comments are crucial to understand the development of the 
verification strategy. %The code has not been optimized for minimal size, nevertheless, the 
%size is very small and runtime is very short for all of the verification.
We can also observe that the verification runtime is also very small: due to the breakdown of internal and external verification, we can verify complex architectures 
within a very short time.  The most effort-consuming step is the design and writing of the verification models, but the actual verification is quick.

\begin{table}[t]
  \centering
  \resizebox{0.9\linewidth}{!}{
  \footnotesize
  \begin{tabular}{ c | c | c | c }
    \hline
    \textbf{Model}        & \textbf{Int. or Ext.}  & \textbf{Lines of Code}    & \textbf{Runtime}  (s)\\ \hline \hline
    VM Startup              & Ext.      & 1159                  & 0.8s \\ \hline
    VM Launch       & Int.        & 462                   & 0.6s \\ \hline
    VM Secure Channel           & Ext.      & 1332                  & 0.3s \\ \hline
    VM Trust Evidence   & Ext.      & 1081                  & 0.2s \\ \hline
    VM Suspend \& Resume      & Ext.      & 1054                  & 0.5s \\ \hline
    VM Mem. Update      & Int.        & 687                   & 0.7s \\ \hline
    VM Terminate      & Int.        & 417                   & 0.8s \\ \hline
  \end{tabular}
  }
  \vspace{-10pt}
  \caption[HyperWall verification evaluation results.]{\small HyperWall verification evaluation results.}
  \label{table:verifProtSummary}

\vspace{-5pt}
\end{table}

\begin{table}[t]
  \centering
  \resizebox{0.9\linewidth}{!}{
  \footnotesize
  \begin{tabular}{ c | c | c | c }
    \hline
    \textbf{Model}        & \textbf{Int. or Ext.}  & \textbf{Lines of Code}    & \textbf{Runtime}  (s)\\ \hline \hline
    External         			& Ext.		& 262               		& 0.2s \\ \hline
    Evidence Collection         		& Int.			& 123               		& 0.1s \\ \hline
    Property Interpretation			& Int.			& 205               		& 0.2s \\ \hline
    Health Checking			& Int.			& 187               		& 0.1s \\ \hline
  \end{tabular}
  }
  \vspace{-10pt}
  \caption[CloudMonatt verification evaluation results]{\small CloudMonatt verification evaluation results.}
  \label{table:verifProtSummary1}
\vspace{-15pt}
\end{table}

\section{Related Work}
\label{sec:related}

%In this section we list many of the available security protocol verification tools, generic model checkers that have been used for security verification, logic languages that focus on protocol security verification and theorem provers.  We also discuss limited work that we could identify on invariant generation.

\bheading{Secure application verification.}
Past work, e.g.,~\cite{Sinha:2015,SuSiLe:17}, has focused 
%Past work offer solutions to verify applications running in secure processors. Sinha et al. \cite{Sinha:2015} proposed a formal method to verify the confidentiality property of enclaves software in SGX architecture. Subramanyan et al. \cite{SuSiLe:17} verified the integrity and secure measurement of enclave software. These work focus 
on verifying software with respect to an ISA. In contrast, we are going one layer below, focusing on the state machines and protocols of the hardware.

\bheading{Secure architecture verification.}
%Past projects have looked at attempting to perform formal verification of an entire architecture \cite{dehon11}. 
Of the different secure architectures, XOM \cite{lie00} and SecVisor \cite{seshadri07} have received the benefit of security verification using model-checking. 
%XOM verification \cite{XOM} checked read-protection and tamper-resistance against a highly simplified model of the instruction set operations on registers and memory, and SecVisor verification \cite{scalableRefMon} used a logical system to reason about how security demonstrated by a small model of the SecVisor reference monitor can scale to the full implementation size. 
In industry, e.g., the IBM 4758 cryptographic co-processor's design included security verification~\cite{smith99}.  These works, however, have not focused on external and internal protocols and interactions as we do.

\bheading{Security verification tools.}
A number of speciality tools exist for security verification and verification of security protocols.
%A number of tools specifically for security verification have been developed. Some tools focus on security protocols. They are designed to automatically verify the cryptographic protocols and find attacks that could undermine the security protocols which focus on secrecy and authentication. 
These tools include HERMES \cite{hermes}, Casrul \cite{casrul}, AVISPA \cite{Armando05}, Scyther \cite{scyther}, ProVerif \cite{proverif}, etc.
%Athena \cite{athena}, etc.
% Brutus \cite{Clarke00}, RVChecker \cite{revere}, STA (Symbolic Trace Analyzer) \cite{Boreale02}, Casper \cite{casper}, etc. 
%
Various model checkers have also been used in security verification. 
%These tools have no built-in security-related features and are meant for functional verification. But they include concepts such as invariants which can be adapted to define security invariants as well. 
Typical model checkers include Maude \cite{maude}, Alloy \cite{jackson2006alloy}, Murphi \cite{Murphi}, CSP \cite{csp}, FDR \cite{fdr}, etc.
Our work does not invent a new tool, rather it shows how architects can leverage existing tools.
%In contrast to the above work, this paper proposes a new and general-purpose verification framework for secure architectures. 
%This framework is independent of verification tools. 
We show that both the protocol verification tools (ProVerif) and model checkers (Murphi) can be used by our framework to implement the security verification task.
We enhanced Murphi with automatic checking for integrity and confidentiality, 
but did not need to make any changes to Proverif.

\bheading{Security verification methodologies.}
As an alternative to modelling, %security verification can work directly with the hardware source code.
projects such as Caisson~\cite{li2011caisson} or SecVerilog~\cite{zhang2015hardware}
work directly with the hardware source code, and
leverage information flow tracking to analyze potential information leaks in an architecture.
Our work does not require hardware source code, and can be complimentary to the approaches that work with HDL code.

\section{Conclusion}
\label{sec:conclusion}

We present a security verification methodology, which is applicable to different security architectures and systems. We break the verification task into external verification and internal verification to achieve scalability of verification. For each type of verification, we propose the methodology for modeling the system and the attackers, deriving security invariants, and creating the implementation. We use two case studies to evaluate our methodology: security verification of a standalone processor architecture, HyperWall, and verification of a distributed cloud system, CloudMonatt. Our case studies show that we can verify the design of complex secure architectures efficiently, discover and fix bugs, and enhance the security of the design. 
%Moreover, they show that design bugs can be discovered, and fixed, thanks to this methodology.
%Security verification can also guide the designers to enhance the security of architectures, and discover potential flaws at the design time. 
We hope that our methodology can be easily adopted by computer architects to
verify the security of their designs, and to do more research in the important area of security verification methodologies.

%Through our definition of a security verification methodology, we are able to verify the important external protocols and internal interactions of an architecture, and complement the functional verification of the mechanisms of trusted hardware with our security checks.  Our methodology for invariant generation allows non-security experts to perform security verification of their hardware architectures, using familiar tools such as Murphi, which are already familiar to them. We showed how to expand the stock Murphi with new integrity, and also confidentiality, invariants.  Murphi contains syntax for expressing invariants, but gives no guidance on what the invariants should be -- one of our contributions is the systematic definition of security invariants for integrity, and also confidentiality. While HyperWall served as our case study, this methodology is applicable to other architectures as well.  Moreover, it has good performance, as discussed below.

\bibliographystyle{ieeetr}
\bibliography{ref}

\end{document}